\documentclass{emulateapj}
\usepackage{graphicx}

\newcommand{\lya}{Ly$\alpha$~}
\def\see{\mbox{$^{\prime\prime}$}}

\shorttitle{Confirmation of galaxies at redshift higher than 7}
\shortauthors{Vanzella et al.}

\begin{document}


\title{Spectroscopic confirmation of two Lyman break galaxies at
 redshift beyond 7}


\author{E. Vanzella,\altaffilmark{1}
L. Pentericci\altaffilmark{2},
A. Fontana\altaffilmark{2},
A. Grazian\altaffilmark{2},
M. Castellano\altaffilmark{2},
K. Boutsia\altaffilmark{2},
S. Cristiani\altaffilmark{1}, 
M. Dickinson\altaffilmark{3},
S. Gallozzi\altaffilmark{2}, 
E. Giallongo\altaffilmark{2},
M. Giavalisco\altaffilmark{4},
R. Maiolino\altaffilmark{2}, 
A. Moorwood\altaffilmark{5},
D. Paris\altaffilmark{2},
and P. Santini\altaffilmark{2}}

\affil{$^{1}$INAF Osservatorio Astronomico di Trieste, Via G.B.Tiepolo 11, 
34131 Trieste, Italy}
\affil{$^{2}$INAF Osservatorio Astronomico di Roma,  Via Frascati 33,00040 
Monteporzio (RM), Italy}
\affil{$^{3}$National Optical Astronomy Observatory, PO Box 26732, Tucson, AZ 85726, USA}
\affil{$^{4}$Department of Astronomy, University of Massachusetts, 710 North Pleasant Street, Amherst, MA 01003}
\affil{$^{5}$European Southern Observatory, Karl-Schwarzschild Strasse, 85748 Garching bei Munchen, German}
\email{vanzella@oats.inaf.it}



\begin{abstract}
  We report the spectroscopic confirmation of two Lyman break galaxies
  at redshift $>$ 7.  The galaxies were observed as part of an
  ultra-deep spectroscopic campaign with FORS2 at the ESO/VLT 
  for the confirmation of $z\simeq7$ ``z--band dropout'' candidates 
  selected from our VLT/Hawk-I imaging survey.  Both galaxies show 
  a prominent emission line at
  9735\AA~and 9858\AA~respectively: the lines have
  fluxes of $\sim 1.6-1.2\times 10^{-17} erg s^{-1}cm^{-2}$ and
  exhibit a sharp decline on the blue side and a tail on the
  red side. The asymmetry is quantitatively comparable to the
  observed asymmetry in $z\sim 6$ Ly$\alpha$ lines, where absorption by 
  neutral hydrogen in the IGM truncates the blue side of the emission 
  line profile. We carefully evaluate the possibility that the galaxies
  are instead at lower redshift and we are observing either 
  [O\,{\sc ii}], [O\,{\sc iii}] or H$\alpha$ emission: however from the
  spectroscopic and the photometric data we conclude that there are no
  other plausible identifications, except for \lya at redshift $>7$,
  implying that these are two of the most robust redshift determination for galaxies
  in the reionization epoch. Based on their redshifts and
  broad--band photometry, we derive limits on the star formation rate
  and on the ultraviolet spectral slopes of the two galaxies. We argue
  that these two galaxies alone are unlikely to have ionized the IGM in
  their surroundings.

\end{abstract}



\section{Introduction}
The recent progress in the exploration of the very early Universe has
been quite remarkable with several groups beginning  to assemble large
samples of candidate high-redshift galaxies ($z >6.5$). These include
narrow-band selected \lya emitters (LAEs) \citep{Ota2010,Ouchi2010,Hu2010}, and 
color-selected Lyman break galaxies (LBGs) from wide and deep surveys
carried out in the near-IR, mainly with WFC3
\citep{Bouwens2010a,McLure2009b} and Hawk-I
\citep[C10a and C10b in the following]{Castellano2010a,Castellano2010b}.

Despite the large number of candidates, spectroscopic
confirmation of a sizable sample of high redshift galaxies is still
lacking.  A systematic spectroscopic follow up is of paramount
importance for three different reasons. First, to evaluate the
accuracy and reliability of the Lyman--break selection technique that, 
albeit quite successful at $z\simeq 6$, has never been validated at these
extreme redshifts. In addition, the intensity of the \lya can
provide clues on the rest frame properties of the observed galaxies: 
knowledge of the exact redshift and the contribution
of \lya and other emission lines to the observed broad--band
magnitudes are needed to derive  
the rest-frame properties of the galaxies from the observed continuum
(SFR, dust content, metallicity). Finally, and more
intriguing, the very visibility of the \lya and the
distribution of its equivalent width in high-z galaxies can provide
useful constrains on the ionization state of the intergalactic
medium (IGM) at epochs less than 800 Myr after the Big-Bang
(e.g. Fontana et al. 2010, F10 in the following; Ouchi et al. 2010).
  
To date, only a few galaxies have been confirmed at $z\simeq 7$ or beyond.
Following  the early
identification  of the $z=6.96$ \lya emitter by \cite{Iye2006},
 a gamma ray burst host was reported at
$z\simeq 8.1$ by both Salvaterra et al. (2009) and Tanvir et al. (2009), but with
considerable uncertainties on the exact redshift, due to the
absence of an emission line in the spectra. Very recently the
detection of a galaxy at z$=$8.6 was claimed by \cite{Lehnert2010},
although the \lya emission line in the spectrum has very low S/N and
a high uncertainty in the flux calibration.

In this context we have started a systematic campaign of
spectroscopic follow-up of $z\simeq7$ ``z--band~dropout'' candidates,
selected from our imaging survey obtained with VLT/Hawk-I
(C10a and C10b).
In F10 we presented the results on the sample selected in the GOODS--S
field.  Out of seven candidates observed, we tentatively detected only
one weak Ly$\alpha$ emission line at $z=6.97$. As discussed in that
paper (see also Stark et al 2010), this very low fraction of
confirmations is at odds with what is expected by extrapolating the
$z=5-6$ surveys which detect a much larger fraction of 
\lya emitters.

Our survey has continued over the two other fields
described in C10b.  In this paper we present the
only two objects with a clear \lya\ line at $z\geq7$ found 
in the entire survey, which are both detected in the BDF4 field 
(Lehnert \& Bremer 2003). The results on the final
sample with a full discussion of their implication is deferred to a
forthcoming paper (Pentericci et al. in preparation).

All magnitudes are in the AB system, and we adopt
$H_0=70$~km/s/Mpc, $\Omega_M=0.3$ and $\Omega_{\Lambda}=0.7$.

\section{Target Selection and Spectroscopic Observations}
The targets were selected according to the criteria described
extensively in C10b: besides the three $z$--band dropout
candidates listed in Table 3 of that paper, other slits were filled
with less secure $z=7$ candidates as well as $i$--band dropouts.

Observations were taken in service mode with the FORS2 spectrograph on
the ESO Very Large Telescope, during July-August 2010. 
We used the 600Z holographic grating, that provides the highest
sensitivity in the range $8000-10000$\AA\  with a spectral resolution
 $R\simeq 1390$ and a sampling of 1.6\AA\  per pixel for a 1\see$\times$12\see~slit. 
The data presented here come from the co-addition of 86 spectra of 665 seconds 
of integration each, on a single mask, for a total of 15.9 hr, with median 
seeing around 0.8\see. 
Series of spectra were taken at two different positions, offset
by 4\see~(16 pixels) in the direction perpendicular to the dispersion.

Standard flat-fielding, bias subtraction and wavelength calibration 
have been applied as in \cite{Vanzella2009} and F10. 
The sky background has been subtracted between 
consecutive exposures, exploiting the fact that the target
spectrum is offset due to dithering.
Before combining frames, 
particular care has been devoted to the possible offset along the wavelength 
direction, by measuring the centroids of the sky lines in the wavelength 
interval 9400-9900\AA. We have also carried out the sky subtraction 
by fitting a polynomial function to the background.
The two approaches provide consistent results.

Finally, spectra were flux-calibrated using the observations of
spectrophotometric standards. Slit losses are small, given the
extremely compact size of the targets and have been neglected 
in the subsequent discussion.

\section{Results}
\subsection{Redshift determination}
We detect a prominent emission line in the spectra of two 
galaxies, candidates 
$BDF-521$ and $BDF-3299$ (C10b), at wavelengths of 9735\AA\ 
and 9858\AA\ respectively. In Figures~\ref{fig1} and ~\ref{fig2} 
we present the sky-subtracted 
extracted 1-dimensional spectra and 2-dimensional spectra 
 for both objects.
No other lines are detected in the rest of the spectra  and no 
continuum is detected for either object.
The total line fluxes  are 1.6$\pm 0.16 \times 10^{-17} erg s^{-1}cm^{-2}$ and  
1.2$\pm0.14 \times 10^{-17} erg s^{1-} cm^{-2}$
respectively. 

Both lines show a clear asymmetric profile with a 
sharp decline on the blue side and a prominent tail on the red side. This
 asymmetric profile, which we attribute to absorption by neutral hydrogen,
 is the best and in many cases unique  diagnostic of high-z \lya
emission.  Indeed most LAEs and LBGs at high redshift are too faint to
 detect the break in the continuum caused by IGM attenuation,
 with the exception of few very 
bright objects \citep{Kodaira2003}.

We  have investigated the possibility that the lines are instead due 
to other  features, such as H$\alpha$, H$\beta$, [O\,{\sc iii}]$\lambda5007$ 
or the doublet [O\,{\sc ii}]$\lambda\lambda3726-3729$ in the spectra of 
lower redshift objects. 
If the detected emission lines were H$\beta$ or 
[O\,{\sc iii}]$\lambda$5007 at lower redshift (z$\sim$1.0 and $\sim$0.95), 
then both emission lines should be seen in the observed wavelength range.
 Actually the other component of the [O\,{\sc iii}] doublet (at $\lambda 4959$) 
should also be observed, although with lower S/N.
No other lines are detected in the spectra. We also  
checked that the positions of the expected lines were not coincident with any bright 
skylines.

In case of H$\alpha$ emission from $z\sim0.50$ galaxies, the
rest-frame EW of the line would be exceeding 300\AA, a value that is
very rarely observed in star forming galaxies at low
redshift \citep{Salzer2005}. Furthermore, we note that these objects
would have a relatively bright continuum also at wavelengths below
1$\mu$, and should therefore be observed in the deep R, I, z--bands,
where instead we set very stringent upper limits for non-detection 
($AB=28-29$ depending on the band, see Table 1 of C10b).

Finally, in the case of [O\,{\sc ii}] emitters, the resolution of our spectra (R=1390) 
 would be enough to distinguish the two components of the doublet, which at $z=1.6$ are
 separated by 8\AA. 
We searched for examples in our masks, and we do detect the [O\,{\sc ii}] doublet 
in two galaxies at z$\sim$1.6, one at z$\sim$1.5 and two at z$\leq$1.3. In particular 
the two [O\,{\sc ii}] at z$\sim$1.6 fall in  the same wavelength region as the lines 
discussed here and have a lower flux compared to our candidates.
In all cases we can clearly distinguish the two components: examples
at redshift 1.3, 1.5 and 1.6 are shown in the inset of Figure~\ref{fig2} 
(top right).
Furthermore, we note that typically the [O\,{\sc ii}] doublet even
  when unresolved produces an opposite asymmetry with respect
  to the \lya shape, because the $\lambda$3726\AA~component 
  is often weaker than the $\lambda$3729\AA~(see also Rhoads et al. 2003).

Since the correct identification of the line is a critical issue when
no other spectral features are present, as mentioned above, the 
asymmetry can be used to distinguish high-z \lya emission  
 from foreground [O\,{\sc ii}], [O\,{\sc iii}], or H$\alpha$ emitters
(see Stern et al. 2000). 
 To test our ability to recognize asymmetric lines with
  the S/N of our spectra, we run Monte Carlo simulations by
  inserting two-dimensional lines of
  different shapes (Gaussian, truncated Gaussian with variable width,
  and [OII] doublet with variable ratio of the two components) in the
  original science frames at positions corresponding to the observed
  wavelengths, normalized to the flux of our lines. 
  The frames have been matched in seeing and spectral resolution and 
  reduced as the real ones.  This was repeated for all slits where we have no
  signal detected.  We find that an asymmetric line is always
  recognized as such if it has a width $> 150 km/s$, and that 
  the [OII] doublet
  in the 1-D spectra is always resolved into two components.

  We also calculated the weighted skewness $S_w$ introduced
  by Shimasaku et al. (2006), to quantify the asymmetry of the two lines.
  \lya emission produces always a positive $S_w$: they
  set a conservative value of $S_w> 3$ to distinguish 
  LAEs from foreground emitters, although there could be also
  LAEs with $S_w<3$.  
  For our two galaxies we find  $S_w > 5-8$ (depending on the 
  wavelength's range assumed for the measurement), very similar to the 
  average  $S_w$ of $z=6.6$ LAEs $\left<S^{z=6.6}_{w}\right>=7.31\pm 1.51$\AA~.



We conclude that both lines can be safely identified with 
\lya emission: this implies redshifts of 7.008$\pm 0.002$ and 7.109$\pm 0.002$
for BDF--521 and BDF--3299, respectively.

\subsection{Continuum fluxes and spectral slopes}

In the following we determine the correct UV continuum magnitudes and
the \lya equivalent widths for each object.

\begin{figure}
\epsscale{1.0}
\plotone{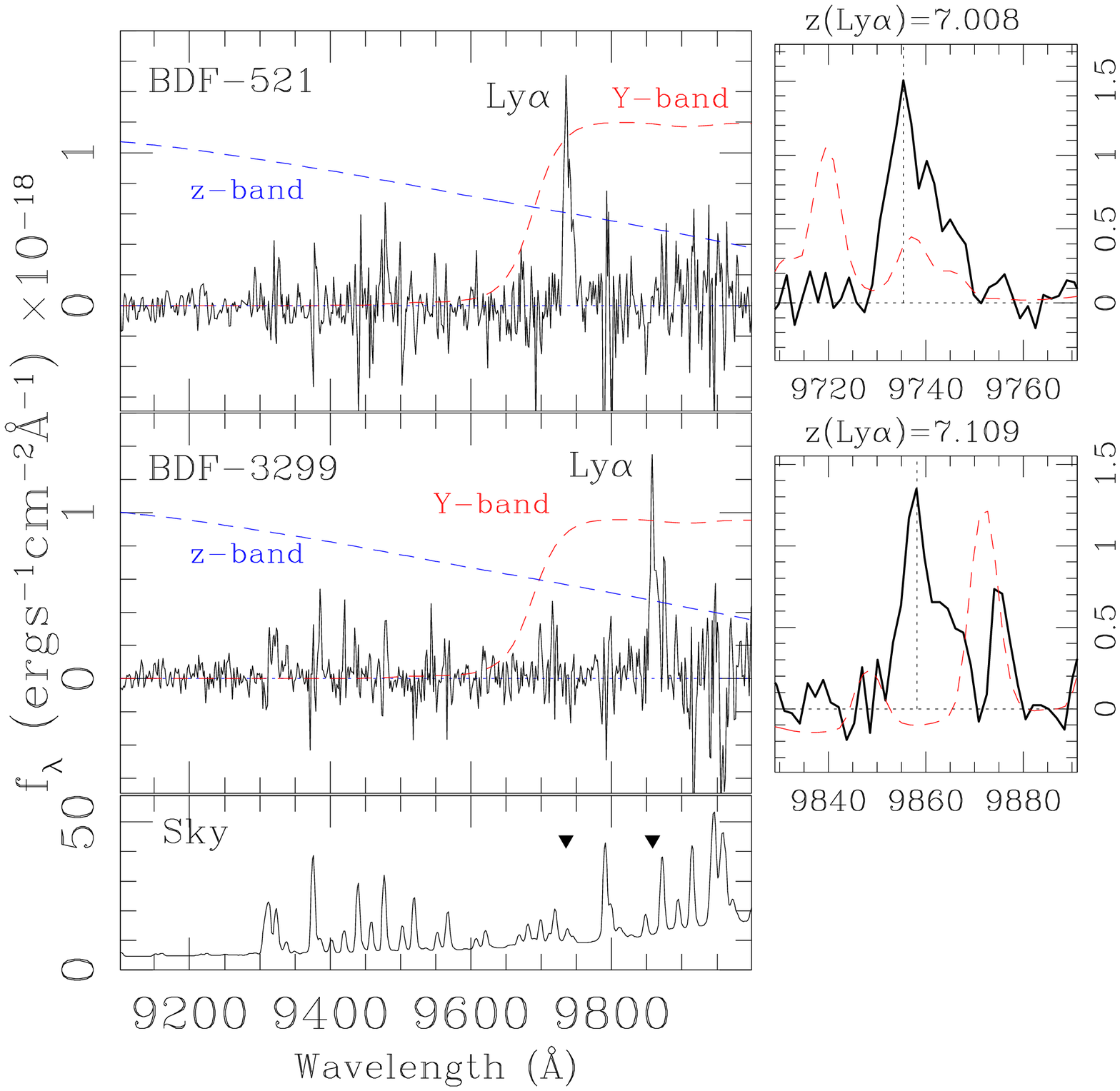}
\caption{1-dimensional spectra of candidates BDF--521 and BDF--3299. On the left
panels spectra are shown with superimposed the z--band and Y--band filters. 
In the bottom the 1-dimensional (flux calibrated) spectrum of the sky is shown
(the position of the \lya lines are marked with triangles).
On the right side the zoomed \lya lines are shown with the position of the peak 
marked with a vertical dotted line. The two lines are significant at around
20 sigma level. In all panels units in the Y-axis are
$10^{-18}~erg~s^{-1}~cm^{-2}$\AA$^{-1}$.
The red dashed lines superimposed to the zoomed \lya (on the right) 
show the spectrum of the sky in arbitrary units. \label{fig1}}
\end{figure}

\begin{figure}
\epsscale{1.0}
\plotone{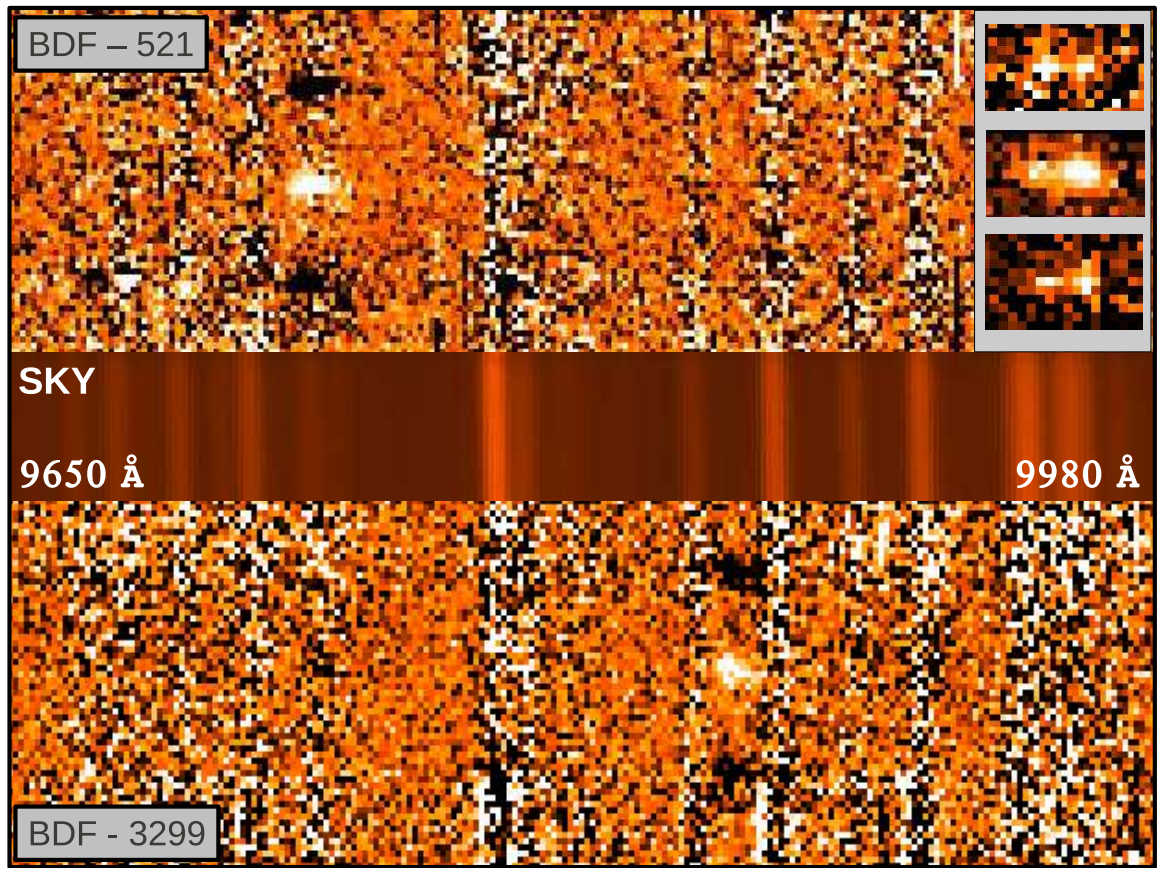}
\caption{2-dimensional sky subtracted spectra of BDF-521 (top) and BDF-3299 (bottom).
In both cases the emission line is clearly detected. The ``negative'' features
above and below the emission (typical of the sky subtraction technique) are also evident.
In the middle the sky spectrum is shown as a reference for sky line positions.
The resolved [O\,{\sc ii}] doublets at redshift 1.3, 1.5 and 1.6 are shown in the 
upper right corner, from bottom to top, respectively. 
\label{fig2}}
\end{figure}

$\bullet$ {\bf BDF--521}: Figure~\ref{fig1} (top panel) shows the
relative position of the \lya line and the two filters $Y$ and
$z$. The IGM opacity affects only 2.7\% of the $Y$--band filter
($<$0.05 mag for a flat $F_{\nu}$ spectrum), therefore we neglect the IGM attenuation.
The observed magnitude is $Y=25.86\pm0.11$, corrected to $Y_{cont}=26.31$ if
the \lya contribution is subtracted. We note that for this source
there is also a 2-sigma detection at mag 28.00$\pm$0.60 in the z-band
which is consistent with the \lya falling at the
edge of the z-band filter. Assuming $Y_{cont}=26.31$ the rest-frame
\lya equivalent width is 64$_{-9}^{+10}$\AA~(where the errors come from the 
line flux uncertainty).

$\bullet$ {\bf BDF--3299}: The observed magnitude is $Y=26.15\pm0.14$,
corrected to $Y_{cont}=26.58$ if \lya is subtracted. However in this case
the IGM attenuation affects  15\% of the $Y$--band filter, absorbing
completely the continuum  blue-ward of the \lya line (see Figure~\ref{fig1}). 
Assuming conservatively a flat spectrum for the continuum, 
we derive $Y_{cont}=26.40$ and a rest-frame \lya equivalent 
width of 50$_{-8}^{+11}$\AA.

Adopting these \lya EWs we derived the slopes of the UV
  continuum, fitting to the broad-band photometry a set of simple
  models with flux scaling as $F_{\lambda}\propto\lambda^{\beta}$, 
which include also the \lya emission and IGM absorption. We derive
  the best fit models and the relevant error analysis with a 
  standard $\chi^2$ minimization.  The results are shown in Figure 3,
  where we plot the best--fitting solutions  along
  with the error analysis (see insets). The best--fit
  slopes are remarkably steep, with $\beta=-3.7$ and $-2.7$ for
  BDF--521 and BDF--3299, respectively,
  similar to those recently claimed for z--band
  dropouts on the basis of new HST WFC3 data ($\beta<-2.5$, Bouwens et
  al. 2010b; Oesch et al. 2010; Bunker et al. 2010; 
  Finkelstein et al. 2010).  However, much
  shallower solutions with $\beta<-1.7$ and $<-0.9$ are still allowed
  at 1-sigma c.l.  ($\Delta\chi^{2}=1$). Clearly, these results are
  mostly constrained by the non-detection in the J and in the K-band
  (see Table 1).


\subsection{Implication for the stellar population}

 We have first explored whether these lines can be ascribed to AGN.
 From the non-detection of NV in the stacked spectrum we derive a 
 1$\sigma$ lower limit on the $Ly\alpha /NV>8$.
This value is higher than the average ratio for QSOs but not unusual
for weak narrow-line AGN found amongst LBGs 
(e.g. Steidel et al. 2002), 
so we cannot  exclude this possibility. However we rely on the fact that
AGN are much rarer than galaxies, and in the following we
will assume that the objects are star forming galaxies.

At the redshifts derived above the implied luminosities for the
\lya lines are 9.3 and 7.3 $\times 10^{42} erg s^{-1}$.
We estimate
the star formation rate (SFR) from the \lya luminosity using the
Kennicutt's relation (Kennicutt 1998) with the case B recombination
theory as $SFR=9.1 \times L(Ly\alpha)$, with L($Ly\alpha$) in units of
$10^{43} erg s^{-1}$. We obtain values around 7--9 $M_\odot yr^{-1}$ (see
Table 1): they represent lower limits since they are
not corrected for absorption effects which depend on various
parameters, including the neutral fraction of the IGM and 
the kinematic status of neutral hydrogen within the galaxies (e.g. Ahn 2004).

From the continuum luminosity we obtain 
$L(UV_{1275})=8.1\times10^{28}$ and $L(UV_{1260})=7.6\times 10^{28}erg s^{-1} Hz^{-1}$.  We
convert luminosities into SFR using the Kennicutt relation for
UV continuum, $SFR(UV)=1.4\times 10^{-28} L(UV)$. 
We determine the flux at 1500\AA~assuming the slope beta derived before.
We obtain SFRs of 8.9 $M_\odot yr^{-1}$ and 9.4 $M_\odot yr^{-1}$ respectively, similar
to those determined from the \lya\ line. These values are obtained
assuming no dust and the agreement of SFR(UV) and SFR($Ly\alpha$) is
consistent with this scenario. Using instead a global fit to the
overall photometry  with Bruzual \& Charlot templates and a Calzetti (2000)
attenuation law, we find best fit values for SFR that are in
excellent agreement with the above determinations, and E(B-V)=0.  
  The standard error analysis on these fits provides the limits one
  can place on the maximum SFR. Given the poor constraints on the J and
  K-bands, we find that for all models acceptable
  at 68\% confidence level $E(B-V)$ is lower than 0.25-0.35, while
  the SFRs upper limits are 160$M_\odot yr^{-1}$ and 
  400$M_\odot yr^{-1}$, respectively, for the two objects, depending on the
  combination of age and metallicity.  Note that the adoption of a
  Gallerani et al. (2010) attenuation curve would provide a SFR limit
  about 2 times lower.

\section{Discussion}

With the spectroscopic identification of these two $z>7$ galaxies we
have moved closer to the epoch of reionization and may possibly be
observing ``re-ionizers'' at work at the end of this process.  While it
is clearly impossible to draw any statistical result out of two
objects only, we discuss here a few implications.

Recently, several authors have argued that the fraction of galaxies
with large Ly$\alpha$ EW increases in the redshift range $4<z<6$
(e.g. Stark et al. 2010; Douglas et al. 2010; V09).  The two emitters
reported here with \lya EW$\ge$50\AA~seem to continue this trend. Indeed the
fraction of z--band dropouts spectroscopically confirmed is much higher
for this field (two out of three candidates observed), compared to the
GOODS-south field (one out of seven candidates, F10).

Therefore the main results of F10, namely that the fraction of
galaxies with \lya EW exceeding 50\AA~seems to show a sharp decline
from $z\sim6$ to $z\sim7$, thus reversing the increasing trend from 
$z\sim4$ to $z\sim6$, appears in conflict with the present observations.

However we anticipate that the result in F10 is
confirmed by the analysis of the final sample that includes a third field
(Pentericci et al. in preparation)
 and a much larger statistics. 
The present spectra indicate that there are considerable
field to field variations and therefore solid conclusions can be 
drawn only from large samples.

Turning to the nature of these objects, one can determine whether
they are able to ionize the IGM around them.  Many models
indeed predict that only intrinsically UV bright galaxies have enough
photons to build reionized ``bubbles'' around them, on reasonably
short timescales \citep{Dayal2008}.  Following Loeb et al. (2005) and
assuming SFR of $10M_{\odot}yr^{-1}$ and age $100Myr$ we derive a
maximum radius of the H\,{\sc ii} region of
$R_{MAX}=0.73(f_{esc})^{1/3}$ Mpc physical with $f_{esc}$ the fraction
of escaping ionizing photons.  We are assuming isolated galaxies
surrounded by a medium that is mostly neutral when they started to
form stars ($z \sim 8$).  In order for the \lya to be transmitted the
optical depth $\tau_{damp}$ to \lya absorption at the galaxy redshift
has to be less than one (Wyithe \& Loeb 2005): this corresponds to a
$R=1.1$ Mpc.  Therefore the radius we have derived would not be large
enough even if a maximum $f_{esc}=1 $ is considered.

As already mentioned the intrinsic SFRs could be considerably higher:  
for the maximum SFR consistent with our photometry, the galaxies could
ionize a large enough region if they had $f_{esc} > 0.1$. Such
values of $f_{esc}$ are not commonly observed in the $z\simeq 3$
Universe with few exceptions, such as the compact LBG at $z=3.8$
recently reported by Vanzella et al. (2010) for which the Lyman
continuum ($\lambda<912$\AA) has been detected directly and whose
non-ionizing features reveals $\beta=-2.1$, very weak interstellar
absorption lines and absence of \lya emission.

We stress that with the current depth of  the near infrared observations,
it is not possible to produce better constraints on the UV slopes, 
dust content and SFR. This issue could easily  be solved with a modest 
investment of HST time, 
by obtaining deep Y, J and H-band WFC3 observations.

If the steep slopes and the dust-free scenarios are confirmed, then the
 galaxies do not have enough photons to ionize the surrounding 
H\,{\sc ii} regions. 
In such a case, the visibility of the \lya line in these two
objects implies the existence of  additional ionizing sources, either
 galaxies and/or objects of different nature. 
It is intriguing to surmise that the field-to-field variations
in \lya visibility we observe indicate that the  
reionization process was very inhomogeneous at these epochs. 
A possible reason is a clustering effect, particularly efficient since
the first objects formed in highly biased regions (e.g. Furlanetto et al. 2006).

We point out that the two confirmed LBGs in the BDF field
are at a physical distance from each other that is  
of the same order as the clustering length of LAEs at $z=6.6$ 
$r_{0}= 2-5 h^{-1}_{100} Mpc$ (Ouchi et al. 2010)
(they are separated by 6$'$ in the plane of the sky corresponding to 1.9 Mpc 
and by 0.101 in redshift space corresponding to a proper distance of 4.4 Mpc). However there is no
evidence of significant enhancement in the density of z--band dropouts in this field.

Another possible mechanism suggested by Wyithe \& Loeb (2005) is
 related  to a previous QSO activity, i.e., the H\,{\sc ii} regions 
generated by quasars 
remain as a fossil after the quasar activity ends, since the recombination 
time is longer than the Hubble time at the mean IGM density 
for z $<$ 8.

These scenarios and their combinations can be tested only when
a sizable sample of high--z sources are confirmed and
their clustering and physical properties are determined.

\acknowledgments Observations were carried out using the Very Large
Telescope at the ESO Paranal Observatory under Programme IDs
085.A-0844, 283.A-5052 and 181.A-0717. 
We thank the anonymous referee for useful comments.
We would like to thank P. Dayal and A. Ferrara for useful comments and 
discussion. 
We acknowledge financial contribution from the agreement ASI-INAF I/009/10/0.




\clearpage

\clearpage

\clearpage

\begin{figure}
\epsscale{0.8}
\plotone{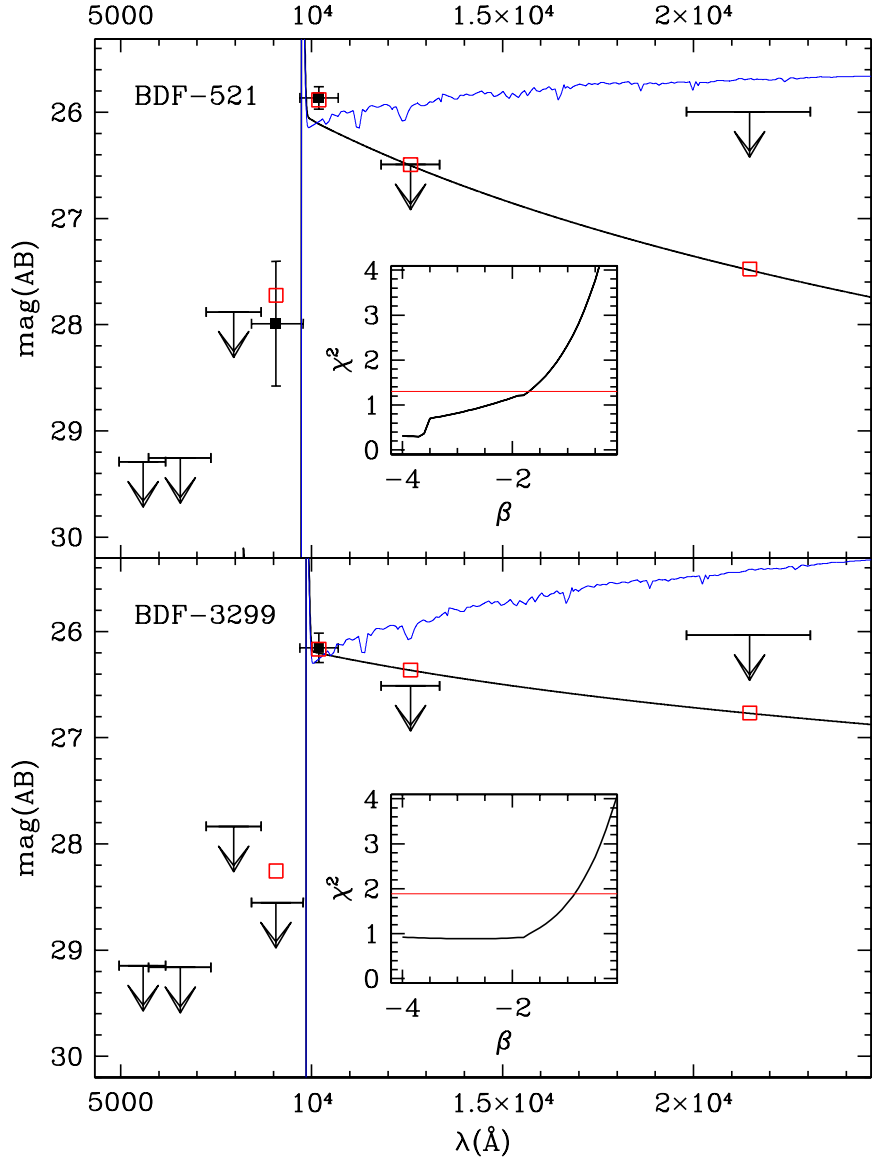}
\caption{
Observed photometry and best-fitting spectral model for BDF-521 (upper panel) and BDF-3299 (lower panel).
Black dots and upper limits correspond to observed magnitudes. The
solid line is a best-fitting $\beta$ model. Red dots are the predicted
magnitudes from the model. The blue thin line is the 
dust-extincted Bruzual \& Charlot model which provides the  maximum SFR with an 
acceptable $\chi^2$. In the inset the $\chi^2$ as a function of $\beta$ is shown 
(the red horizontal line marks the $\Delta \chi^2=1$).
\label{fig3}}
\end{figure}

\clearpage
\begin{table}
\scriptsize
\begin{center}
\caption{Spectroscopic and photometric properties of the two redshift seven galaxies\label{tbl-2}}
\begin{tabular}{cccccccccccc}
\tableline\tableline
ID & RA, DEC&Redshift & f($Ly\alpha$) & SFR(\lya)       & $EW_{rest}$& FWHM\tablenotemark{a}  & $S_w$ & z &Y&J & K  \\
   & J2000  &         &         & $M_\odot yr^{-1}$& \AA       & $km s^{-1}$ & \AA  &    & &1$\sigma$&1$\sigma$ \\
\tableline
\\
BDF-521  & 336.9444, -35.1188 & 7.008$\pm$0.002 & 1.62$\pm$0.16 & 8.5 & 64 & 240 &$8.2^{+2}_{-1}$ & 28.00 &25.86 & $>$26.5  & $>$26.0    \\
BDF-3299 & 337.0511, -35.1665 & 7.109$\pm$0.002 & 1.21$\pm$0.14 & 6.6 & 50 & 200 &$>5.9\tablenotemark{b}$ &$>$28.55  &26.15 & $>$26.5  & $>$26.0    \\
\tableline
\end{tabular}
\tablecomments{f($Ly\alpha$) in units of $10^{-17}ergs^{-1}cm^{-2}$\AA$^{-1}$. 
(a) Measured directly from the line and corrected for instrumental broadening. 
(b) This is considered a lower limit since the red side of the emission line is 
truncated by the presence of a bright skyline (see Figure 1).}
\end{center}
\end{table}


\begin{thebibliography}{}
\expandafter\ifx\csname natexlab\endcsname\relax\def\natexlab#1{#1}\fi

\bibitem[Ahn (2004)]{Ahn}
  Ahn, S.-H., 2004, \apj, 601, 25

\bibitem[{{Bouwens} {et~al.}(2010{\natexlab{a}}){Bouwens}, {Illingworth},
  {Oesch}, {Stiavelli}, {van Dokkum}, {Trenti}, {Magee}, {Labb{\'e}}, {Franx},
  {Carollo}, \& {Gonzalez}}]{Bouwens2010a}
{Bouwens}, R.~J., {Illingworth}, G.~D., {Oesch}, P.~A., {et~al.}
  2010{\natexlab{a}}, \apjl, 709, L133

\bibitem[{{Bouwens} {et~al.}(2010{\natexlab{b}}){Bouwens}, {Illingworth},
  {Oesch}, {Trenti}, {Stiavelli}, {Carollo}, {Franx}, {van Dokkum},
  {Labb{\'e}}, \& {Magee}}]{Bouwens2010b}
{Bouwens}, R.~J., {Illingworth}, G.~D., {Oesch}, P.~A., {et~al.}
  2010{\natexlab{b}}, \apjl, 708, L69



\bibitem[Bunker et al. (2010)]{Bunker2010}
 Bunker, A., J., Wilkins, S., Ellis, R.~S., Stark, D.~P., Lorenzoni, S., 
 Chiu, K., Lacy, M., Jarvis, M.~J., Hickey, S., 2010, MNRAS, 409, 855


\bibitem[Calzetti et al. (2000)]{calzetti2000}
Calzetti, Daniela, Armus, Lee, Bohlin, Ralph C., Kinney, Anne L., 
Koornneef, Jan, Storchi-Bergmann, Thaisa, 2000, \apj, 533, 682

\bibitem[{{Castellano} {et~al.}(2010a){Castellano}, {Fontana}, {Boutsia},
  {Grazian}, {Pentericci}, {Bouwens}, {Dickinson}, {Giavalisco}, {Santini},
  {Cristiani}, {Fiore}, {Gallozzi}, {Giallongo}, {Maiolino}, {Mannucci},
  {Menci}, {Moorwood}, {Nonino}, {Paris}, {Renzini}, {Rosati}, {Salimbeni},
  {Testa}, \& {Vanzella}}]{Castellano2010a}
{Castellano}, M., {Fontana}, A., {Boutsia}, K., {et~al.} 2010, \aap, 511, A20+ (C10)

\bibitem[Castellano et al.(2010b)]{Castellano2010b} 
Castellano, M., Fontana, A., Paris, D., Grazian, A., Pentericci, L.,
Boutsia, K., Santini, P., Testa, V., Dickinson, M., Giavalisco, M.,  2010, A\&A,
524, 28 (C10b)

\bibitem[{{Dayal} {et~al.}(2008){Dayal}, {Ferrara}, \& {Gallerani}}]{Dayal2008}
{Dayal}, P., {Ferrara}, A., \& {Gallerani}, S. 2008, \mnras, 389, 1683


\bibitem[{{Douglas} {et~al.}(2010){Douglas}, {Bremer}, {Lehnert}, {Stanway}, \&
  {Milvang-Jensen}}]{Douglas2010}
{Douglas}, L.~S., {Bremer}, M.~N., {Lehnert}, M.~D., {Stanway}, E.~R., \&
  {Milvang-Jensen}, B., 2010, MNRAS, 409, 1155


\bibitem[{{Fontana} {et~al.} (2010)}]{Fontana2010}
Fontana, A., Vanzella, E., Pentericci, L., et al.\ 2010, \apj, 725, 205



\bibitem[Finkelstein et al. (2010)]{finkelstein2010}
Finkelstein, S., L., Papovich, C., Giavalisco, M., Reddy, N., A., 
Ferguson, H. C., Koekemoer, A. M., Dickinson, M., 2010, \apj, 719, 1250

\bibitem[Furlanetto et al. (2006)]{Furlanetto2006}
Furlanetto, S.~R., Zaldarriaga, M., Hernquist, L., 2006, MNRAS, 365, 1012

\bibitem[Gallerani et al. (2010)]{gallerani2010}
Gallerani, S., Maiolino, R., Juarez, Y., Nagao, et al., 2010, A\&A, 523, 85



\bibitem[{{Hu} {et~al.}(2010){Hu}, {Cowie}, {Barger}, {Capak}, {Kakazu}, \&
  {Trouille}}]{Hu2010}
{Hu}, E.~M., {Cowie}, L.~L., {Barger}, A.~J., {et~al.}, 2010, \apj, 725, 394

\bibitem[{{Iye} {et~al.}(2006){Iye}, {Ota}, {Kashikawa}, {Furusawa},
  {Hashimoto}, {Hattori}, {Matsuda}, {Morokuma}, {Ouchi}, \&
  {Shimasaku}}]{Iye2006}
{Iye}, M., {Ota}, K., {Kashikawa}, N., {et~al.} 2006, \nat, 443, 186


\bibitem[Kennicutt (1998)]{kenni98}
 Kennicutt, Robert C., Jr., 1998, \apj, 498, 541

\bibitem[Kodaira et al. (2003)]{Kodaira2003} 
  Kodaira, K., Taniguchi, Y., Kashikawa, N., Kaifu, N., 
  Ando, H., Karoji, H., Ajiki, M., Akiyama, M., Aoki, K.,
  et al., 2003, PASJ, 55, 17

\bibitem[Lehnert 
\& Bremer(2003)]{Lenhert2003} Lehnert, M.~D., \& Bremer, M.\ 2003, \apj, 593, 630 
\bibitem[Lehnert et al.(2010)]{Lehnert2010} Lehnert, M.~D., et al.\ 
2010, \nat, 467, 940 

\bibitem[Loeb et al. (2005)]{Loeb2005}
  Loeb, A., Barkana, R., Hernquist, L., 2005, \apj, 620, 553



\bibitem[{{McLure} {et~al.}(2010){McLure}, {Dunlop}, {Cirasuolo}, {Koekemoer},
  {Sabbi}, {Stark}, {Targett}, \& {Ellis}}]{McLure2009b}
{McLure}, R.~J., {Dunlop}, J.~S., {Cirasuolo}, M., {et~al.} 2010, \mnras, 403,
  960

\bibitem[{{Oesch} {et~al.}(2010){Oesch}, {Bouwens}, {Illingworth}, {Carollo},
  {Franx}, {Labb{\'e}}, {Magee}, {Stiavelli}, {Trenti}, \& {van
  Dokkum}}]{Oesch2009b}
{Oesch}, P.~A., {Bouwens}, R.~J., {Illingworth}, G.~D., {et~al.} 2010, \apjl,
  709, L16

\bibitem[{{Ota} {et~al.}(2010){Ota}, {Iye}, {Kashikawa}, {Shimasaku}, {Ouchi},
  {Totani}, {Kobayashi}, {Nagashima}, {Harayama}, {Kodaka}, {Morokuma},
  {Furusawa}, {Tajitsu}, \& {Hattori}}]{Ota2010}
{Ota}, K., {Iye}, M., {Kashikawa}, N., {et~al.}, 2010, \apj, 722, 803


\bibitem[{{Ouchi} {et~al.}(2010){Ouchi}, {Shimasaku}, {Furusawa}, {SAITO},
  {Yoshida}, {Akiyama}, {Ono}, {Yamada}, {Ota}, {Kashikawa}, {Iye}, {Kodama},
  {Okamura}, {Simpson}, \& {Yoshida}}]{Ouchi2010}
{Ouchi}, M., {Shimasaku}, K., {Furusawa}, H., {et~al.}, 2010, \apj, 723, 869



\bibitem[Rhoads et al.(2003)]{Rhoads2003} Rhoads, J.~E., et al.\ 2003, \aj, 125, 1006 

\bibitem[Salzer et al. (2005)]{Salzer2005} Salzer, J.~J., Jangren, 
A., Gronwall, C., Werk, J.~K., Chomiuk, L.~B., Caperton, K.~A., Melbourne, 
J., \& McKinstry, K.\ 2005, \aj, 130, 2584 

\bibitem[{{Salvaterra} {et~al.}(2009){Salvaterra}, {Della Valle}, {Campana},
  {Chincarini}, {Covino}, {D'Avanzo}, {Fern{\'a}ndez-Soto}, {Guidorzi},
  {Mannucci}, {Margutti}, {Th{\"o}ne}, {Antonelli}, {Barthelmy}, {de Pasquale},
  {D'Elia}, {Fiore}, {Fugazza}, {Hunt}, {Maiorano}, {Marinoni}, {Marshall},
  {Molinari}, {Nousek}, {Pian}, {Racusin}, {Stella}, {Amati}, {Andreuzzi},
  {Cusumano}, {Fenimore}, {Ferrero}, {Giommi}, {Guetta}, {Holland}, {Hurley},
  {Israel}, {Mao}, {Markwardt}, {Masetti}, {Pagani}, {Palazzi}, {Palmer},
  {Piranomonte}, {Tagliaferri}, \& {Testa}}]{Salvaterra2009}
{Salvaterra}, R., {Della Valle}, M., {Campana}, S., {et~al.} 2009, \nat, 461,
  1258



\bibitem[Shimasaku et al. (2006)]{Shimasaku2006}
 Shimasaku, K., Kashikawa, N., Doi, M., Ly, C., Malkan, M.~A., Matsuda, Y., 
 Ouchi, M., Hayashino, T., Iye, M., et al., 2006, PASJ, 58, 313


\bibitem[{{Stark} {et~al.}(2010){Stark}, {Ellis}, {Chiu}, {Ouchi}, \&
  {Bunker}}]{Stark2010}
{Stark}, D.~P., {Ellis}, R.~S., {Chiu}, K., {Ouchi}, M., \& {Bunker}, A.,
  2010, MNRAS, 408, 1628

\bibitem[Steidel et al. (2002)]{Steidel2002}
  Steidel, C.~C., Hunt, M.~P., Shapley, A.~E., Adelberger, K.~L., 
  Pettini, M., Dickinson, M., Giavalisco, M., 2002, \apj, 576, 653

\bibitem[Stern et al. (2000)]{stern2000}
  Stern, D., Bunker, A., Spinrad, H., Dey, A., 2000, \apj, 537, 73

\bibitem[{{Tanvir} {et~al.}(2009){Tanvir}, {Fox}, {Levan}, {Berger},
  {Wiersema}, {Fynbo}, {Cucchiara}, {Kr{\"u}hler}, {Gehrels}, {Bloom},
  {Greiner}, {Evans}, {Rol}, {Olivares}, {Hjorth}, {Jakobsson}, {Farihi},
  {Willingale}, {Starling}, {Cenko}, {Perley}, {Maund}, {Duke}, {Wijers},
  {Adamson}, {Allan}, {Bremer}, {Burrows}, {Castro-Tirado}, {Cavanagh}, {de
  Ugarte Postigo}, {Dopita}, {Fatkhullin}, {Fruchter}, {Foley}, {Gorosabel},
  {Kennea}, {Kerr}, {Klose}, {Krimm}, {Komarova}, {Kulkarni}, {Moskvitin},
  {Mundell}, {Naylor}, {Page}, {Penprase}, {Perri}, {Podsiadlowski}, {Roth},
  {Rutledge}, {Sakamoto}, {Schady}, {Schmidt}, {Soderberg}, {Sollerman},
  {Stephens}, {Stratta}, {Ukwatta}, {Watson}, {Westra}, {Wold}, \&
  {Wolf}}]{Tanvir2009}
{Tanvir}, N.~R., {Fox}, D.~B., {Levan}, A.~J., {et~al.} 2009, \nat, 461, 1254

\bibitem[{{Vanzella} {et~al.}(2009){Vanzella}, {Giavalisco}, {Dickinson},
  {Cristiani}, {Nonino}, {Kuntschner}, {Popesso}, {Rosati}, {Renzini}, {Stern},
  {Cesarsky}, {Ferguson}, \& {Fosbury}}]{Vanzella2009}
{Vanzella}, E., {Giavalisco}, M., {Dickinson}, M., {et~al.} 2009, \apj, 695,
  1163 (V09)

\bibitem[Vanzella et al. (2010)]{vanzella2010}
  Vanzella, E., Giavalisco, M., Inoue, A.~K., Nonino, M., et al., 2010, \apj, 725, 1011

\bibitem[Wyithe \& Loeb (2005)]{WL2005}
  Wyithe, J.~S.~B. \& Loeb, A., 2005, \apj, 625, 1

\end{thebibliography}
\end{document}